# Carrier Transport in Magnesium Diboride: Role of Nano-Inclusions

A.M. Awasthi[1*], S. Bhardwaj[1], V.P.S. Awana[2], A. Figini Albisetti[3], G. Giunchi[3], and A.V. Narlikar[1]

[1]UGC-DAE Consortium for Scientific Research, University Campus, Khandwa Road, Indore- 452 001, India.
[2]National Physical Laboratory, Dr. K.S. Krishnan Marg, New Delhi- 110 012, India.
[3]EDISON S.p.A., R&D Dept., Foro Buonaparte 31, I-20121 Milano, Italy.

## ABSTRACT

Anisotropic-gap and two-band effects smear out the superconducting transition ($T_c$) in literature reported thermal conductivity of $MgB_2$, where large electronic contributions also suppress anomaly-manifestation in their negligible phononic-parts. Present thermal transport results on scarcely-explored specimens featuring nano-inclusions exhibit a small but clear $T_c$-signature; traced to relatively appreciable phononic conduction, and its dominant electronic-scattering. The self-formed MgO as extended defects strongly scatter the charge carriers, and minutely the phonons with their longer-mean-free-path near $T_c$. Conversely, near room temperature, the shorter-dominant-wavelength phonon's transport is hugely affected by these nanoparticles; undergoing ballistic to diffusive crossover, and eventually entering the Ioffe-Regel mobility threshold regime.

---

[*] E-mail: amawasthi@csr.res.in

Soon after the discovery of superconductivity (SC) in MgB$_2$,[1] intense research sought to enhance its critical current density ($J_c$).[2] Incorporation of the pinning centers dramatically increases $J_c$, by hindering the current-dissipative dynamics of the flux-vortices. As the functional formations of the SC alloy include wires and tapes, pure single crystals are unsuitable for this purpose; more so because of their little flux-pinning. While the grain-boundaries and other structural defects in polycrystals do enhance pinning, they invariably broaden the transition and reduce the $T_c$.[3] In MgB$_2$, self-formed/introduced nanoparticles have, however, proved optimal for the $J_c$-improvement, without seriously sacrificing the SC characteristics.[4-5] The multiband SC-character in this alloy is manifested in the heat capacity $C_p \sim T^\alpha$ ($\alpha \approx 2$) below $T_c$, with its non-linear $\boldsymbol{B}$-dependence,[6] and in scanning tunneling spectroscopy (STS).[7] Its Fermi surface has been calculated to consist of 2-D cylindrical sheets and 3-D tubular networks;[8] serving as two distinct electronic bands taking part in the carrier transport.[9] In the SC state the hole-type $\sigma$-bands contribute to the larger gap $\Delta_\sigma$ ($\approx$ 7meV) and the electron type $\pi$-bands lead to smaller gap $\Delta_\pi$ ($\approx$ 2meV), both vanishing at the bulk $T_c$.[7] Static lattice disorder proves ineffective for the equalization of SC-gap amplitudes in MgB$_2$, as these bands' disparity forbids appreciable elastic inter-band scattering.[9]

Generally, thermal conductivity $\kappa(T)$ is expected to show either a shoulder or a peak/kink anomaly right below $T_c$, depending respectively on whether the heat transport is dominated by the fermions or the lattice, and a more complex behavior otherwise/with anisotropic gap structure.[10] Intriguingly, however, heat transport studies on a wide variety of MgB$_2$ specimens (well-conducting, with residual resistivity $\rho_0 \leq 50\mu\Omega$cm) in the form of single crystals[11] and undoped[12-15] or doped[3] polycrystals have so far failed to exhibit any anomaly at their SC-transition. For $\kappa(T)$ to depict the SC-signature in double-gapped MgB$_2$, the following criticalities compel themselves to prevail across the transition. (A) Necessarily elastic (and large) scattering of fermions, to allow accurate evaluation of the electronic-contribution ($\kappa_{el}$) to thermal conductivity by (then) strictly applicable Wiedemann-Franz (W-F) law (equality of fermions' electrical and thermal relaxation times; $\tau_{fer}^{el} \cong \tau_{fer}^{th}$),[16] resulting in a (collateral to subdued *el*-transport) *relatively higher* phononic-contribution ($\kappa_{ph}$) to the heat transport.

(B) Necessarily $\Delta_\sigma$–driven sharp drop in the quasiparticles' number below $T_c$, to abruptly affect the (necessarily fermionic, minutely off-defects) scattering of the phonons ($\tau_{ph}^{-1} = [\tau_{ph}^{fer}]^{-1} + [\tau_{ph}^{def}]^{-1} \approx [\tau_{ph}^{fer}]^{-1}$). Our MgB$_2$ polycrystal having self-formed MgO nano-inclusions realizes these criteria; revealing the long-eluded $T_c$-anomaly in its thermal conductivity.

Polycrystalline MgB$_2$ samples were prepared by first pelletizing the well-mixed (*Reidel-de-Haen*, assay 99%) Mg and (*Fluka*, assay 95–97%) amorphous B powder, along with 2-3wt% SiC/Diamond nano-additives, and encapsulating in an alumina boat placed inside a soft Fe-tube. The charge was heated at 800 °C for 2 hrs, and then cooled to room temperature in the Argon gas flow at ambient pressure. For optimized synthesis and functionality details see e.g.[5] X-ray diffraction (XRD) was recorded with a Rigaku diffractometer, scanning electron microscopy (SEM) was carried out with a Leo-440 (Oxford) instrument, and transmission electron microscopy (TEM) was performed using Tecnie G2-20. Resistivity was measured by standard four-probe technique. High-sensitivity thermal conductivity data were obtained on a sintered pellet by conventional steady-state method, after the specimen was surface-cleaned and polished. Temperature gradients across the sample thickness were measured via thermocouple using Keithley-182 voltmeter. Sample's base temperatures ($T \pm 5$mK) were maintained using Lakeshore DRC-93CA controller, in nominal steps of (min.) 0.5K (below 15K) up to (max.) 10K (above 170K). Repeated runs at shifted temperatures confirmed the data-reproducivity, and provided higher density over 5-12K and 33-40K.

Scherrer's analysis of the resolvable peak in XRD (fig1 inset a), of naturally-formed MgO nano-inclusions provides their median size (~12nm) and SEM (fig1 inset b) shows 3–6μm grains of MgB$_2$. Grain-inhomogeneity and porous-voids seen in the TEM results[5] subtly distinguish our specimen-- the grain boundary consists of $d_{GB}$ ~ 3-4nm thick amorphous layers (upper inset c), and a relatively pore-free grain-section confirms dispersed (7-15nm) spheroid MgO nano-inclusions (lower inset c). SiC/Diamond nano-additives were undetected. The resistivity $\rho(T)$ plot of our polycrystalline MgB$_2$ is shown in fig.2 (main panel), with the critical temperature $T_c = 37$K, residual resistivity $\rho_0 \approx 160$ μΩcm, and the residual resistivity ratio (RRR = $\rho_{300K}/\rho_0$) of 2.94. We have also shown (fig.2 inset a) our measured resistivity of an optimal MgB$_2$ specimen with homogeneous microstructure ($\rho_0 = 2.67$ $\mu\Omega$cm, RRR = 5.37, SEM shown as sub-inset a), prepared by the liquid-Mg infiltration of Boron powders,[17] and similar to the sample labeled B in ref.[18] Highly phase-pure polycrystalline (pc) samples have been reported with $\rho_0 \sim 0.4$ $\mu\Omega$cm and[19] RRR $\sim$ 20. Here, we illustrate an effective conducting cross-sectional area analysis, and point out its inefficacy for the present case. Defined by Rowell;[19] the empirical ratio [$F = (\rho_{300K}-\rho_0)/8.5$ $\mu\Omega$cm] of the $T$-dependent resistivity $\Delta\rho$ to that of the MgB$_2$ polycrystals with almost-connected grains $\Delta\rho_{pc}^{min}$ is taken to indicate the grain-boundary (i.e., extrinsic) effects; $1/F$ being the current-carrying fractional-area of the sample. While $\Delta\rho_{pc}^{min}$ is close to the single crystal (sc)[20] value $\Delta\rho_{sc} = 4.3$ $\mu\Omega$cm, our huge $\Delta\rho = 310.7$ $\mu\Omega$cm implies very low grain-connectivity ($1/F = 1/36 = \Delta\rho_{pc}^{min}/\Delta\rho = 2.73\%$); apparently indicating large inter-grain effects at the face value! However, our microstructural details call for a different perspective, as follows.

Considering one grain (of size-scale $d_G$ and sectional-area $A_G \propto d_G^2$) plus its boundary (of size-scale $d_{GB}$ and sectional-area $A_{GB} \propto [d_G^2 d_{GB}]^{2/3}$) as the repeating motif, conducting (CND) and insulating (INS) cross-sectional areas configure as: $A_{pores}+A_{MgO}+A_{CND}=A_G$, $A_{pores}+A_{MgO}+A_{GB}=A_{INS}$, and $A_{INS}/A_{CND}=F-1$, thus $A_{INS}/A_{GB}=(1+A_G/A_{GB})\times(1-F^{-1})$. For our specimen, the share of grain-boundary to the insulating cross-section thus reckons as only a few percent ($A_{GB}/A_{INS} \approx (d_{GB}/d_G)^{2/3}/(1-F^{-1}) \sim 10^{-2}$); the non-MgB$_2$ (i.e., the insulating) area is mostly provided by the (intra-grain) nano-sized MgO-inclusions and pores! Defining a metric of the extrinsicity, based on the same assumption implied in the $F$-calculation (i.e., $[\Delta\rho - \Delta\rho_{PC}^{min}] \propto \rho_{GB}$, with a universal prefactor); for our practical sample, the "fractional extrinsic contribution" ($\eta = \rho_{GB}/\rho_0$) to the total elastic resistivity figures as only 56% higher than that for our optimally intrinsic specimen ($T_c$ = 38K, $1/F_{opt}$ = 73.5%, fig.2 inset a). The grain-boundary (amorphous, $d_{GB}$ < 4nm) must then consist of ~25Å thick layers of BO$_x$, bordered by the oxygen-rich MgB$_2$.[19] Such thin grain-boundaries contribute only scarcely to $T_c$ -reduction & -width as SC-insulating/normal-SC (S-I/N-S) Josephson junctions below $T_c$, and in enlarging the residual $\rho_0$. The MgO nano-inclusions here as extended defects cause strong intra-grain elastic scattering and subdue the electronic mean free path $\Lambda_{el}$ (~ 1nm) to below the superconducting coherence length $\xi_{clean}$ (~ 5nm). This effectively realizes the circumstance of a dirty superconductor, where the high concentration ($n$) of point defects causes $\Lambda_{el} \sim (n)^{-1/3}$. These figures signify the bulk transport in our sample with nano-inclusions as neo-intrinsic, which is not simply resolvable into separate intra/inter-grain parts; rendering the usual interpretation of the *F*-parameter inapplicable here.

Purely for the sake of parametric estimation, we write our $\rho(T)$ as the sum of elastic (intra-grain/grain-boundary convolute) and inelastic (thermal) terms; $\rho(T) = \rho_0 + \rho_{ph}(T)$. We take the Bloch-Grüneisen[16] (B-G) $\rho_{ph}(T) = (m-1)\rho'\Theta_D \left(\frac{T}{\Theta_D}\right)^m J_m\left(\frac{T}{\Theta_D}\right)$ with $J_m\left(\frac{T}{\Theta_D}\right) = \int_0^{\Theta_D/T} \frac{x^m dx}{(e^x - 1)(1 - e^{-x})}$ to represent the *T*-dependent part. Here $\Theta_D$ is the Debye temperature, $\rho'$ is the high-temperature resistivity-slope/temperature-coefficient of resistivity (TCR), and $m$ = 3 to 5. In the limiting case $T < 0.1\Theta_D$, we can write $\rho(T) = \rho_0 + AT^m$ (*A* is constant). Our resistivity at low-*T* is fittable using this power-law with $m$ = 3 (typical of the transition metals)[11,19] and residual resistivity $\rho_0 \approx 160$ μΩcm (fig.2); fit deviating from the data above ≅ 140K. A few telescopic iterations for the B-G $\rho_{ph}(T)$ converge to provide $\Theta_D$ = 1100K, agreeing with the majority,[6,13] and the yet largest of TCR $\rho'$ = 1.64 μΩcm/K.[9,12] From the modified 'Mooij-plot'[21] of literature-compiled values (fig.2 inset b), our fitted $\rho'$ comes across as rather logical extrapolation of its empirical correlation with $\rho_0$; [$\text{Ln}(\rho'/\rho'_{min}) = 0.156\{\text{Ln}(\rho_0/\rho_0^{min})\}^{1.626}$]. The middle point (closed circle) was determined using the data of ref.;[14] by B-G fitting their $\rho(T)$ with $\Theta_D$ = 915K and $m$ = 2.3 (non-integral *m* reported e.g., ref.[3,15]). Regression of $\rho'$ with $\rho_0$ and the allied[12] $\rho' >> \rho'_\sigma = 0.39$ μΩcm/K indicate either strong *el-ph*-coupling effects[22] or a violation of the Mathiessen's rule, in our naturally mesocomposite alloy. Large (intra-grain) TCR complies with the σ-bands-governed *el*-transport.

Figure 3 shows our thermal conductivity data from 300K down to 6K, with absolute values about half an order of magnitude smaller than the lowest reported[5,12] and, with a clearly discernible small peak right below $T_c$ (inset b, left *y*-scale), similar to that seen in the high temperature superconductor $Bi_2Sr_2CaCu_2O_8$.[23] Without degrading the $T_c$-characteristics in resistivity, the peculiar intra-granular/inhomogeneity effects carry over to reduce $\kappa_{el}$ (evaluated below); creating the favorable circumstance for the $T_c$-manifestation in thermal-transport data, evidently not realized neatly by the bulk-doping.[3,11] Resultant appreciable phononic-contribution ($\kappa_{ph}$) has ensured that the $T_c$-signature survives the smearing. We evaluate the electronic-contribution, assuming the Wiedemann-Franz (W-F) law to apply (more appropriate for highly-resistive specimens, and insensitive to its inter/intra-grain details).[16] $\kappa_{el}(T) = L_0T/\rho \cong \kappa(T)/2$ compares well with the reported ones that feature a broad maximum[14] ($L_0 = 2.45 \times 10^{-8}$ W$\Omega$ K$^{-2}$ is the Lorentz number).[16] Consequent to the highly-scattered phonons losing their extended character, the lattice contribution $\kappa_{ph}(T) = \kappa(T) - \kappa_{el}(T)$ rises steadily (similar to the generic feature in disordered solids), along with the absence of three-phonon anharmonic (Umklapp) processes (which cause the high-$T$ decrease in $\kappa_{ph}(T)$ of relatively cleaner/single-crystal samples).[3,11]

Granting the validity of the W-F law to the normal electrons, the upper (lower) bound contribution in the SC state to the total thermal conductivity $\kappa_s(T)$, of the quasiparticles (QP) excited across the $\Delta_\pi$ ($\Delta_\sigma$) SC-gap can be obtained by considering only their intraband elastic scattering.[9] Thus, based on the Bardeen-Rickayzen-Tewordt two-fluid formulation,[24] the ratio of superconducting to normal electronic-contribution to thermal conductivity is

$$\left.\frac{\kappa_{es}(T)}{\kappa_{en}(T)}\right|_{\pi,\sigma} = \left.\frac{2F_1(-y) + 2y\ln(1+e^{-y}) + y^2/(1+e^y)}{2F_1(0)}\right|_{\pi,\sigma} \quad (1),$$

where $y = \Delta_{\pi,\sigma}(T)/k_BT$ and $F_j(-y) = \int_0^\infty \frac{z^j dz}{1+e^{z+y}}$. Using the two band-gap values[7] as $\Delta_\pi(0) = 2.3$meV and $\Delta_\sigma(0) = 7.1$meV, the above ratios calculated for the $\pi$- and $\sigma$- bands are shown in fig.3 inset a. For proceeding to evaluate the lattice contribution, we consider only the $\sigma$-band, as already established to be applicable to such poorly-conducting samples. The W-F thermal conductivity of normal electrons at low temperature being $\kappa_{en} = L_0T/\rho_0$, their actual Bardeen-Cooper-Schrieffer (BCS) quasiparticle contribution $\kappa_{\sigma s}(T)$ is shown as thick solid curve in the main panel of fig.3. In particular, the $T$-dependence of $\kappa_{el}(T)$ is rather continuous across the $T_c$, which explains for most cases reported, the smearing-out of the SC-transition in their $\kappa_{tot} \cong \kappa_{el}$.[12] The smoothening is more in the purer specimens, due to the increased band-averaging (and closeness to $\kappa_{tot}$) of $\kappa_{el}$.[13] Expectedly, phononic conductivity [$\kappa_{ph}(T) = \kappa_{tot}(T) - \kappa_{el}(T)$] features the SC-anomaly more prominently than does the measured $\kappa_{tot}$.

Note that $\kappa_{el}^{W-F}$ fully bears the influence of grain-inhomogeneity and grain-boundaries; since (as shown below) at low $T$'s, both the mean free path ($\ell_{ph}$) and thermally equivalent wavelength ($\lambda_{ph}^{eq}$) of phonons are larger than the size-scale ($d_{ins}$) of these insulating regions. Therefore, $\sigma$-band transport should dominate in samples having up to a *quarter* of our specimen's resistivity (approx. $\rho'/4\rho'_\sigma \cong 1.05 \geq 1$ as per fig.2 inset b, along with $4/F \cong 0.11$ still << 1 for the 'fully-connected' and homogeneous-grained 'ideal' polycrystal). Maintenance of the $T$-dependence of the quasiparticle-number (and hence of the phonons'-scattering by them) then ensures the sameness of the $\kappa_{ph}$-contribution (as-evaluated above for our specimen). Combining the latter as such with correspondingly *quadrupled* electronic thermal conductivity of a *simulated* specimen alone strips a thus-*simulated* $\kappa_{tot}^{sim}$ of a clear $T_c$-kink anomaly (leaving only a remnant sub-$T_c$ hump, fig.3 inset b, right $y$-axis). The cutoff ($\kappa_{el}^{sim}/\kappa_{tot}^{sim} \cong 80\%$ at $T_c$) thus forbids the clear SC-manifestation for the higher (fractional) *el*-contributions. Our optimal sample too does not show any SC-signature in $\kappa(T)$ (fig.3 inset c), though the reasons here are a large contribution of the $\pi$-band (up to $x_\pi^{opt} \cong 70\%$; $\rho'_\pi \leq \rho'_{opt} = 6.65E-2 << \rho'_\sigma$)[12-13] to $\kappa_{el}$ and the latter's lion's-share ($\kappa_{el}/\kappa_{tot} \geq 90\%$) of the heat-transport.

Using Peierls-Boltzmann transport equation and harmonic lattice degrees of freedom,[16] we write

$$\kappa_{ph} = \frac{1}{3} C_p^{ph} \upsilon_s \ell_{ph} = \frac{1}{3} \ell_{ph} \left(\frac{\upsilon_\parallel + 2\upsilon_\perp}{3}\right) \frac{9pR}{V_M} \left(\frac{T}{\Theta_D}\right)^3 \int_0^{\Theta_D/T} \frac{x^4 e^x dx}{(e^x - 1)^2} \quad (2),$$

(here $p = 3$ and $\upsilon_s$ = sound velocity), the evaluated phonon mean free path $\ell_{ph}$ is shown in fig.4 (left $y$-axis), together with thermally-equivalent phonon wavelength[25] (top $x$-axis) $\lambda_{ph}^{eq} = \upsilon_s/\omega^{eq} = \hbar\upsilon_s/k_B T$, and their characteristic ratio $\lambda_{ph}^{eq}/\ell_{ph}$ (right $y$-axis). Besides noticing the wide (four decades') variation in $\ell_{ph}$, we observe remarkable features here, not directly obvious from thermal conductivity data. Firstly, $\ell_{ph}$(6K) ~ 1.3μm is very similar to its values reported e.g., by Schneider et al.[15] Over the narrow 6-10K range, there are almost no quasiparticles (QP), and the point defects are ineffective as scatterers (dominant phonon wavelength $\lambda_{ph}^{eq} >> n_{imp}^{-3}$); still,[3] grain-boundaries, sheet-like faults, and strain fields of dislocations scatter the higher energy (shorter wavelength) vibrational modes until below 4K (say), boundary scattering is expected to square off $\ell_{ph}$ to its upper bound (viz, the grain size $d_G$ ~ 3-6μm, ref.[5]). At yet lower temperatures, the phonon thermal conductivity $\kappa_{ph}(T)$ is expected to track the lattice heat capacity $C_{ph}$ ~ $T^3$, as e.g. reported in refs.[11,15] The QP-depleted regime below 10K ($\ell_{ph}$ ~ $T^{-1.39}$) is clearly discernible from the one above (10K < $T$ < 37K), where the QP's decide the electronic ($\kappa_{es}$) and affect the phononic ($\kappa_{ps}$) thermal conductivities. As the *ph-ph* scattering is ruled out at small $T/\Theta_D$ ~ $O(10^{-2})$, subtracting the (extrapolated) low-$T$ ('no QP's) part ($\propto T^{1.39}$) from the total scattering rate ($\Omega_{ph} = \upsilon_s / \ell_{ph}$) of the phonons, we estimate the rate of phononic scattering off the 'fermions' ($\Omega_{ph}^{fer}$) near $T_c$, as shown in fig.4 inset a. The abrupt jump in the power-law behavior of $\Omega_{ph}^{fer}(T)$ across $T_c$ (from ~ $T^{4.19}$ to ~ $T^{2.35}$, along with 20nm $\geq \ell_{ph}^{fer} << \ell_{ph}^{imp} \geq$ 100nm) demonstrates fermions as the major scatterers ($\ell_{ph} / \ell_{ph}^{fer}$ > 84%) of the phononic carriers over this regime. Here, the nano-inclusions/pores and grain-boundaries (the insulating regions) heavily subduing the *el*-transport are less-effective in scattering the phonons ($\ell_{ph}$ ~ 200Å > $d_{ins}$ ~ 40-100Å); precisely fulfilling the criteria described earlier to observe the superconducting-anomaly in the MgB$_2$ thermal transport.

In the normal state, above ≅ 50K phonon's thermally equivalent wavelength ($\lambda_{ph}^{eq}$) and mean free path ($\ell_{ph}$) both fall below 10nm, precisely the size-scale of the nano-inclusions, which dictate the phonon transport onwards. Co-significantly, the ratio $\lambda_{ph}^{eq} / \ell_{ph}$ exceeds 1 which, in thermal transport parlance signifies the 'incipient phonon localization'.[25-26] Here, phonon's ballistic attribute is no longer retained and, instead of the specular-collision type, its scattering turns to more of a (still coherent) wave nature. Thus, due to both reflection and transmission at each scattering event, the phonon heat transport becomes diffusive and less efficient. We find $\ell_{ph} \sim T^{-2}$ over this regime (fig.4 inset b), where ($T < \Theta_D/20$) $C_{ph} \sim T^3$ gives $\kappa_{ph} \sim T$, tracking the $T$-linear $\kappa_{el}$ (fig.3). Another break in the phonon-behavior is witnessed across 70K (more so in the ratio $\lambda_{ph}^{eq}/\ell_{ph}$), above which the inelastic (thermal) scattering of electrons becomes appreciable (the sub-linearity of $\kappa_{el}(T)$, with corresponding faster decrease of $\ell_{ph}$, fig.3). The ensuing local plateau in $\kappa_{ph}(T)$ has been suggested as related to the 'weak localization' of phonons;[25] compensation of 'stronger' decrease of $\ell_{ph}(T)$ and the local increase of $C_{ph}(T)$ with the rising temperature, as per eq.(2). Above 120K, $C_{ph}(T)$ rises relatively faster vis-à-vis the now tailing off $\ell_{ph}(T)$, with $\kappa_{ph}(T)$ increasing again. Above 180K, the super-linear increase of $\rho(T)$ makes $\kappa_{el}(T)$ drop, and the leveling-off of the $\ell_{ph}$ (post-max. $\lambda_{ph}^{eq}/\ell_{ph} \cong 2\pi$, fig.4) now causes $\kappa_{ph}(T)$ to track $C_{ph}(T)$ [over this regime $2\pi \ell_{ph}/\lambda_{ph}^{eq} \sim 1$ and $\ell_{ph}(T) \sim T^{-1} \Rightarrow \ell_{ph}(\omega_{eq}) \sim \omega_{eq}^{-1}$].[26-27] Interestingly, $\ell_{ph}$ settles to its minimum value ~ 3Å ($\equiv$ lattice constant $a$, ref.[5]) near the room temperature--- known as the 'phononic Ioffe-Regel (I-R) crossover', signifying lattice-vibrations' confinement to unit-cell dimensions. Consequent to this, integrity of the individual phonons is obscured (the 'wave-vector' $k_{ph} = 2\pi/\lambda_{ph}^{eq}$ physically irrelevant), as it scatters incoherently; the lattice heat transport now assumes a random-walk like hopping character.[27-28]

In summary, the elusive superconductivity signature registers clearly in thermal transport studies of our MgB$_2$ polycrystal; MgO nano-inclusions as extended defects make irrelevant the otherwise hindering effects of two-bands, gap-anisotropy, and the anomaly cancel-off. Significantly, our twin criteria explain both the presence of $T_c$-kink in the measured $\kappa(T)$ of our specimen, plus its absence in an optimal-specimen's data and in a simulation. Above $T_c$, strong scattering of phonons off the nano-inclusions localizes the atomic vibrations, realizing lattice heat transport generic of the amorphous systems. The electronic-transport in nano-inclusioned MgB$_2$ seems akin to the single-gap ($\Delta_\sigma$) BCS kind in its superconducting state. Spectral study of the contended SC-anomaly criteria, and of the *el-ph* coupling effects[22] are pertinent to pursue further.

**Acknowledgements**

AMA thanks R. Rawat for providing the resistivity data, D.M. Phase for the SEM, and N.P. Lalla for the TEM measurements. Dr. P. Chaddah is acknowledged for scientific support and encouragement.

**Figure Captions**

1. **MgB$_2$ structure/microstructure.** Main panel shows the XRD pattern along with inset (a) enlarged MgO peak, fitted for grain-size (12.2nm) estimate. (b) SEM micrograph shows 3-6μm grains (10μm scale shown lower-left). (c) Upper-half: TEM pictures clearly reveal the 3-4nm thick (most probably) amorphous BO$_x$ inter-grain region (20nm scale shown lower-right), with selected area electron diffraction (SAED) of the MgB$_2$ grain; Lower-half: intra-granular spheroidal 7-15nm MgO nano-inclusions (50nm scale shown lower-right), with their SAED.

2. **MgB$_2$ resistivity.** Cubic power-law and the Bloch-Grüneisen (B-G) fits to the normal state data are also shown. Left inset a: resistivity of an optimal MgB$_2$ specimen (similar to sample B, ref.[18]), along with its SEM micrograph (30μm scale shown lower-left in the sub-inset). Right inset b: empirical regression of the residual resistivity and the high-$T$ resistivity-slope of typical specimens; closed triangle, square, and diamond (all ref.[12]), closed circle[14], closed star (present).

3. **MgB$_2$ thermal conductivity** and its electronic and lattice composition. Contrast the "continuous" electronic behavior across the $T_c$ vis-à-vis rather distinct phononic contributions. Inset (a): different $T$-dependences of $\kappa_{es}/\kappa_{en}$ for the individual $\pi$- and $\sigma$- bands ($\sigma$-band carried through the final analysis). Inset (b) contrasts the clear $T_c$-kink of the present work with its absence in a simulated data. Inset (c) shows our $\kappa(T)$ data on an optimal sample,[18] with no discernible $T_c$-anomaly.

4. **MgB$_2$ phononic length scales.** Temperature dependences of phonon mean free path $\ell_{ph}$ (left y-axis) and the characteristic ratio $(\lambda^{eq}/\ell)_{ph}$ (right y-axis), highlighting the various regimes of dominant scatterers and phonon-localization. Top x-axis: $\lambda^{eq}_{ph} = \upsilon_s/\omega^{eq} = \hbar\upsilon_s/k_BT$. Inset (a) unambiguously marks the sharp change at $T_c$ in the power-law behavior of phonons' scattering rate off the fermionic quasiparticles, while (b) depicts the $\ell_{ph} \sim T^{-2}$ 'incipient phonon localization' regime. Clearly benchmarked are the ballistic/diffusive and diffusive/confined phononic crossovers, not explicit in $\kappa_{ph}(T)$. The $\kappa_{ph}$-plateau relates to the weak localization of phonons.[25] Vibrational Ioffe-Regel regime manifests near room temperature, as the $\ell_{ph} \sim$ lattice-constant roll-off.[26]

**Fig.1**

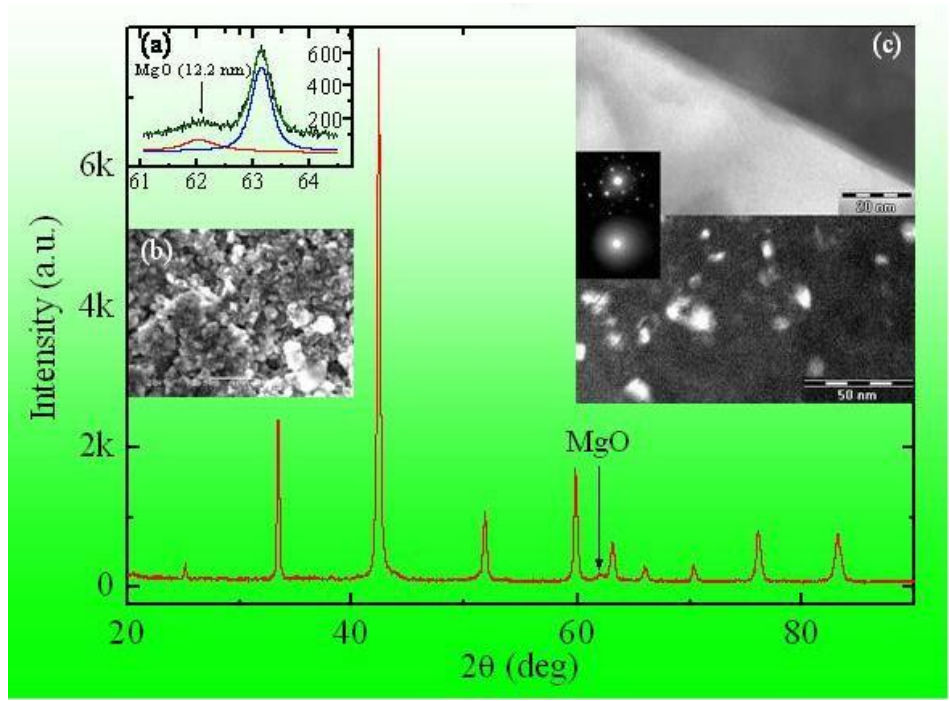

**Fig.2**

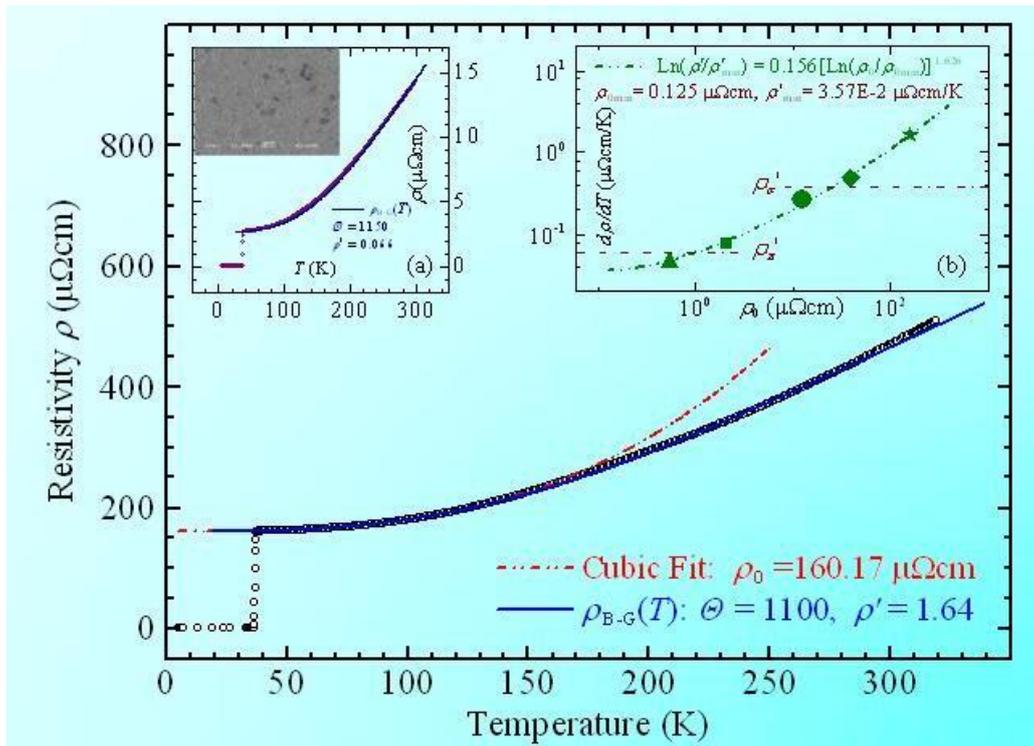

**Fig.3**

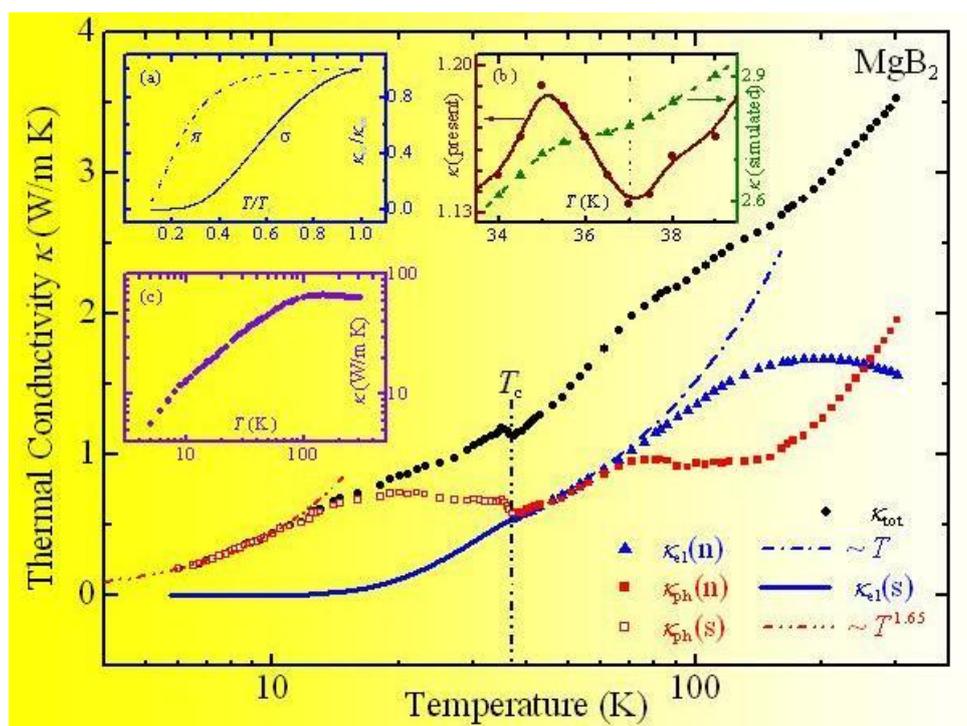

**Fig.4**

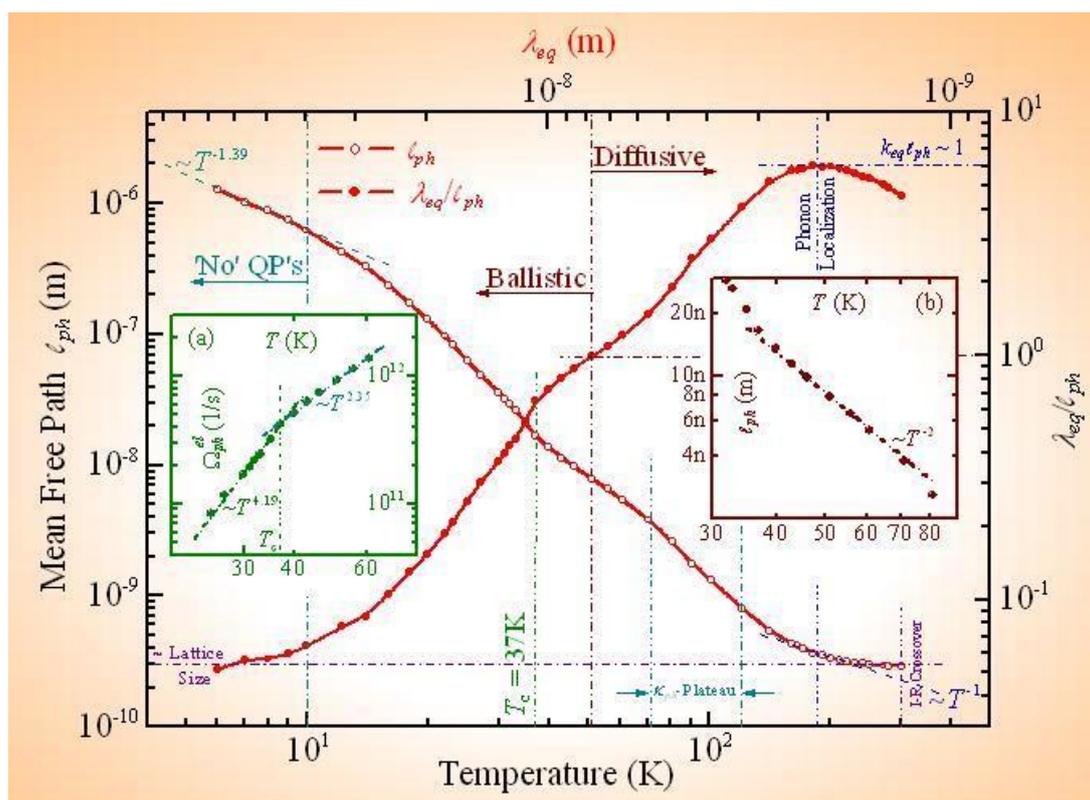